# Weaving Attention U-net: A Novel Hybrid CNN and Attention-based Method for Organs-at-risk Segmentation in Head and Neck CT Images

Zhuangzhuang Zhang, Tianyu Zhao, Hiram Gay, Weixiong Zhang, Baozhou Sun

Washington University in St. Louis, MO, USA

**Purpose:** In radiotherapy planning, manual contouring is labor-intensive and time-consuming. Accurate and robust automated segmentation models improve the efficiency and treatment outcome. We aim to develop a novel hybrid deep learning approach, combining convolutional neural networks (CNNs) and the self-attention mechanism, for rapid and accurate multi-organ segmentation on head and neck computed tomography (CT) images.

**Methods:** Head and neck CT images with manual contours of 115 patients were retrospectively collected and used. We set the training/validation/testing ratio to 81/9/25 and used the 10-fold cross-validation strategy to select the best model parameters. The proposed hybrid model segmented ten organs-at-risk (OARs) altogether for each case. The performance of the model was evaluated by three metrics, i.e., the Dice Similarity Coefficient (DSC), Hausdorff distance 95% (HD95), and mean surface distance (MSD). We also tested the performance of the model on the Head and Neck 2015 challenge dataset and compared it against several state-of-the-art automated segmentation algorithms.

**Results:** The proposed method generated contours that closely resemble the ground truth for ten OARs. On the Head and Neck 2015 challenge dataset, the DSC scores of these OARs were $0.91 \pm 0.02$, $0.73 \pm 0.10$, $0.95 \pm 0.03$, $0.76 \pm 0.08$, $0.79 \pm 0.05$, $0.87 \pm 0.05$, $0.86 \pm 0.08$, $0.87 \pm 0.03$, and $0.87 \pm 0.07$ for brain stem, chiasm, mandible, left/right optic nerve, left/right submandibular, and left/right parotid, respectively. Our results of the new Weaving Attention U-net demonstrate superior or similar performance on the segmentation of head and neck CT images.

**Conclusions:** We developed a deep learning approach that integrates the merits of CNNs and the self-attention mechanism. The proposed weaving attention U-net (WAU-net) can efficiently capture local and global dependencies and achieves state-of-the-art performance on the head and neck multi-organ segmentation task.

## 1. INTRODUCTION

Head and neck cancer (HNC), the seventh most common malignancy, attracts significant research interests worldwide.[1] Radiotherapy treatment (RT) is typically offered to around 75% of all HNC patients[2], which also brings acute and prolonged effects such as mucositis, skin desquamation, loss or alteration of taste, fibrosis, trismus, edema, xerostomia, dental decay, and soft tissue and bone necrosis.[3] Current clinical practices utilize advanced RT techniques, such as intensity-modulated RT (IMRT), to improve target dose conformality and organs-at-risk (OARs) sparing, relying on accurate OAR delineations of computed tomography (CT) images.[4] HNC treatments sometimes are subject to rapidly changing target (tumor growth or shrinkage) and OARs volumes (weight loss resulting in decreased reproducibility and higher doses, salivary gland shrinkage[5]), so treatment plans need to be adjusted accordingly. The replanning requires transferring old contours and adjusting accordingly or a new round of manual contouring.[1] However, manual delineation is labor-intensive and time-consuming and may also be inconsistent because physicians work slice by slice.[6] HNC IMRT cases are especially arduous with more and smaller OARs; manual contouring of each HCN IMRT case takes an average of 2.7 hour[7], rendering it laborious and inefficient for adaptive treatment. Besides, manual contouring introduces considerable inter and intro-observer variations into the delineation results due to various sizes and shapes of targets and OARs, which potentially compromises treatment outcomes. To the rescue, rapid and accurate automated segmentation methods can substantially speed up the process and provide consistent and reliable delineations, presenting promising clinical value, especially in low resource settings.

Automated OARs segmentation on head and neck (HN) CT images has been extensively studied, yielding numerous models for the task. Before machine learning revolutionized medical image semantic segmentation, popular approaches for the problem are model-based, atlas-based, or both.[8] Model-based methods learn from ground-truth segmentation masks to build statistical models capturing the shape and appearance of segmentation objects, which will be fitted with active shape/appearance models.[9,10] In contrast, atlas-based models register multiple atlases CTs to the target CT, propagating atlas contours, during which pixel-wise label predictions are generated with statistical label fusion, elastic transformation, or other fitting models.[11-14] There are also models combining the merits of these two general approaches.[15,16]

Machine learning rapidly grew in the past decade with outstanding generalization and feature representation abilities.[17] Compared to traditional approaches, these learning-based methods no longer rely on deformable registration accuracy or require atlas-to-target CT registration.[1] Early machine learning segmentation models used Markov random field (MRF)[18], supported vector machine[19], and random forest[20]. However, modern machine learning segmentation models mostly used deep learning[1,21-24]. Deep learning segmentation methods outperform early machine learning models due to their excellent abilities of feature extraction, representation, and generalization. Larger deep learning models contain abundant trainable parameters, resulting in higher learning potential and training costs.[25] Convolutional neural networks (CNNs) are the most commonly adopted architecture for image processing.[25] One major advantage of CNNs is their ability to capture local relations with small filters and propagate short-range dependencies by stacking multiple layers.[26] The first CNN-based segmentation method, fully convolutional network (FCN), proposes to replace fully connected layers with convolutional layers, extending the model function from image classification to semantic segmentation.[27] Stemming from FCN, various CNN-based segmentation architectures have been proposed for different applications, among which U-net[28] is initially designed for medical image segmentation. U-net modifies the up-sampling path of FCN and composes a "U" shape structure with its symmetrical encoder and decoder paths[28]. The U-net architecture has prevailed in the medical image segmentation field and many variations have been developed, including U-net++[29], V-net[30], and Deep attention U-net (DAU-net)[31].

The existing deep learning HN multi-organ segmentation models often use CNN as their backbones[32], adopt a U-net style architecture[33,34], and boost the performance with diverse techniques:

- **Multi-stage models**[21,22,35,36] have been proposed to divide the segmentation task into several sub-tasks. Wang et al.[22] propose a two-stage segmentation approach that uses two 3D U-net networks, one for OARs localization and the other for segmentation. Yang et al.[21] propose to link three networks for slice identification, region of interest (ROI) localization, and segmentation. A recursive method also proposes to locate and crop the ROIs on brain CTs with three stages.[35] Guo et al.[36] propose a stratified organ at risk segmentation framework that executes on three parallel branches targeting anchor segmentation, mid-level segmentation, and small & hard detection.
- **Multi-modality models** combine the advantages of CT and Magnetic resonance imaging (MRI) to achieve better segmentation results. Liu et al.[1] use generative adversarial networks (GANs)[37] to synthesize MRI (sMR). They combine the soft-tissue information (sMR) and bony structure information (CT) for better segmentation outcomes.[1]

Although the existing CNN-based methods perform well on HN multi-organ segmentation, an intrinsic weakness of convolutional layers hinders the model performance. CNNs capture local short-range pixel-to-pixel dependencies with small filters and propagate short-range dependencies across the whole image. However, global relations are not explicitly modeled in this propagation process. Moreover, the dependency propagation is achieved by stacking filters (i.e., two $3 \times 3$ filters have a receptive field of $5 \times 5$), resulting in very deep models with a large number of parameters to train, high computation costs, and memory requirements.[38]

The self-attention mechanism[39,40] can adequately model long-range dependencies. It is initially proposed in a transformer model for natural language processing, featuring its ability to model relations between all pairs of word tokens regardless of their positions.[39] The idea has been adopted in the computer vision (CV) field by treating patches/pixels of the image as tokens.[41] Transformer-based models, such as Medical Transformer[38] and Pyramid Medical Transformer[42], perform better than traditional CNN-based methods on gland and nuclei segmentation datasets. However, the potential for hybrid models for HN multi-organ segmentation, combining the merits of CNNs and self-attention mechanism, has not been fully explored.

We propose a novel hybrid method, namely weaving attention U-net (WAU-net), to take advantage of the power of feature extraction of CNNs and the power of self-attention mechanism. The proposed WAU-net adopts the structure of U-net++[29] and integrates with self-attention layers. It uses CNNs to extract low-level features and axial attention blocks[43,44] to model global relations at multiple levels of the architecture efficiently. Within one forward propagation, the model is able to segment ten OARs, i.e., the brain stem, chiasm, mandible, spinal cord, left/right optic nerve, left/right parotid, and left/right submandibular. We test the model performance on two datasets: an in-house dataset with 115 HNC subjects and the Head and Neck 2015 challenge dataset. We evaluate the model with three metrics, i.e., Dice Similarity Coefficient (DSC), Hausdorff distance 95% (HD95), and mean surface distance (MSD). The proposed method generates delineation masks that closely resemble ground truths. The quantitative results from the new model are compared to the results of the state-of-the-art methods.

## 2. METHODS AND MATERIALS

### 2.A. Datasets

Planning CT and structures for 115 HNC patients were retrospectively collected and approved by the Institutional review board (IRB) of Washington University in St. Louis to form our in-house dataset. All CT images were acquired using a 16-slice CT scanner with an 85-cm bore size (Philips Brilliance Big Bore, Cleveland, OH, US). The CT images were acquired with a $512 \times 512$ matrix and 1.5 mm slice thickness. Each patient has 150–200 slices. The ten OARs to be delineated include brain stem, chiasm, mandible, spinal cord, optic nerve (left and right), submandibular (left and right), and parotid (left and right). OARs' contours, serving as the ground truth, were drawn by two radiation oncologists with over 10-yr experience, and consensus contours were generated using the Eclipse treatment planning system (Varian Medical Systems, CA). We held out 25 cases for testing and conducted the 10-fold cross-validation on the rest 90 cases.

We also used the Head and Neck 2015 Challenge data[8] for model evaluation. The dataset provides 33 training cases and 15 testing cases, each of which has ground truth delineations for 9 OARs: brainstem, chiasm, mandible, optic nerve (left and right), submandibular (left and right), and parotid (left and right).

### 2.B. Weaving attention U-net overview

The network (Figure 1) is woven with convolutional layers and self-attention layers, hence the name "weaving attention U-net." It takes in CT images and generates one-hot encoded label maps. The input image is converted to feature maps by convolution blocks. Subsequently, the feature maps are down-sampled and convoluted and then up-sampled and fused with previous feature maps at the same level. After reaching the bottom of the encoder path, meaning we have feature maps 11-15 ready, the feature maps are passed through self-attention blocks at their corresponding levels (Figure 1). Finally, starting at the lowest level, the feature maps (except the ones on the lowest level) are fused with the lower-level feature maps and then up-sampled to restore the original resolution. One final convolution layer is used to generate one-hot encoded label maps. We used the cross-entropy loss function for training.

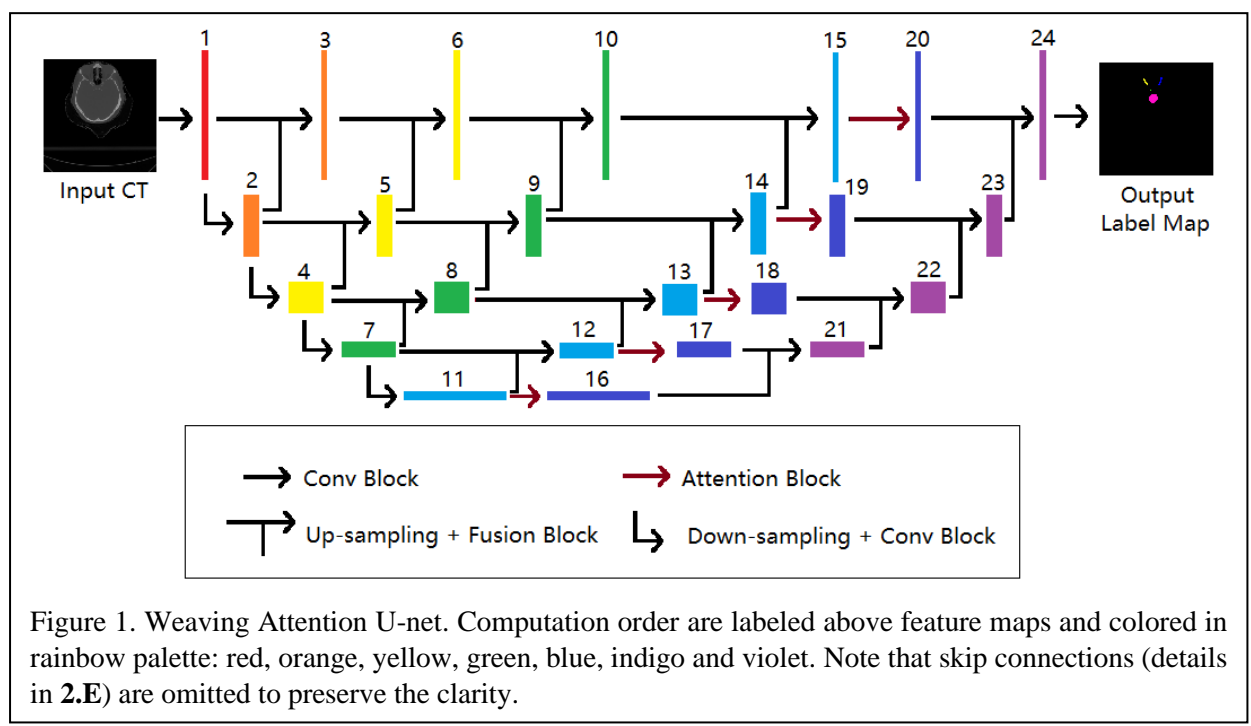

Figure 1. Weaving Attention U-net. Computation order are labeled above feature maps and colored in rainbow palette: red, orange, yellow, green, blue, indigo and violet. Note that skip connections (details in **2.E**) are omitted to preserve the clarity.

Compared with the original U-net[28], which only uses a simple skip connection between its encoder path and decoder path, we added dense convolutional layers and self-attention layers to better pass semantics from the encoder path to the decoder path. The idea was inspired by U-net++[29], and our major innovation is the integration of multi-level self-attention mechanisms. Each up-sampling and fusion block also combines previous feature maps at the same level (details in **2.E**).

### 2.C. CNN blocks

We stack three CNN layers for each CNN block, each of which is $3 \times 3$ with stride-1 and zero paddings. The numbers of filters at different levels are [64, 128, 256, 512, 1024]. We use stride-2 max-pooling layers with a kernel size of $2 \times 2$ for down-sampling and stride-2 deconvolution layers for up-sampling. The final convolution layer is $1 \times 1$ with stride-1, and the number of output channels matches the number of OARs plus one for the background.

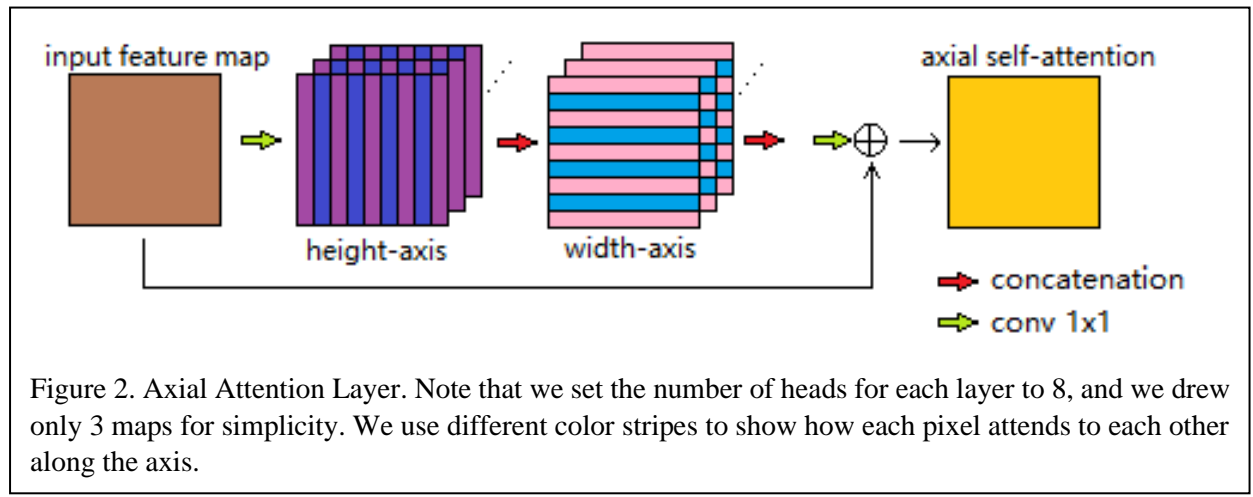

Figure 2. Axial Attention Layer. Note that we set the number of heads for each layer to 8, and we drew only 3 maps for simplicity. We use different color stripes to show how each pixel attends to each other along the axis.

### 2.D. Multi-level axial attention

The self-attention mechanism[39] is designed to explicitly model all pairs of relations between all pairs of word tokens in a sentence regardless of their positions. In computer vision Transformers, image pixels or patches are treated as tokens for the self-attention mechanism.[41] Intuitively, each token is mapped to a key, a value, and a query. Each token uses its query to interact with other tokens' keys, and their matching degrees are the weights with which it aggregates other tokens' values. Mathematically, consider an input image or feature map $X \in H \times W \times C$, where $H$ is height, $W$ width, and $C$ the number of channels, the self-attention map $Y$ is computed as:

$$Q = W_q X, \quad K = W_k X, \quad V = W_v X, \tag{1}$$

$$Y = \mathrm{softmax}\left(\frac{QK^T}{\sqrt{d_k}}\right)V, \tag{2}$$

where $Q$, $K$, and $V$ are queries, keys, and values computed by weight matrix $W_q$, $W_k$, and $W_v$, respectively; $d_k$ is the number of dimensions of the keys and queries; $T$ is the matrix transposing symbol.[39] The matching degree of queries and keys is computed as $\mathrm{softmax}\left(\frac{QK^T}{\sqrt{d_k}}\right)$ in Eq. 2.

Explicitly modeling all pairs of token-to-token relations can effectively capture dependencies of any range, and they are equally to be learned by the model. On the contrary, CNNs aggregate local information within their filter regions and model long-range dependencies by using large filters or stacking multiple small filters, resulting in large models and unsatisfactory model performance. We want to use attention layers with different scales to compensate for such weakness of CNNs. However, assigning weights to all pairs of the relations creates quadratic computation complexity. For a feature map of $X \in H \times W \times C$, there are $HW$ tokens and $(HW)^2$ pairs of relations, which are not scalable for high-resolution images. We choose to use axial attention[43] for approximation. Axial attention[43,44] applies self-attention along the height and width axes of the image separately (Figure 2). For each token, the axial attention computation first aggregates information from all tokens in the same column and then from all tokens in the same row. While computing along the width axis, each token indirectly attends to all other tokens globally. Axial attention reduced the computation complexity from $(HW)^2$ to $HW \times (H + W)$ for $HW$ tokens. Mathematically, the axial attention mechanism for an input feature map $X \in R^{H \times W \times C_{in}}$ with positional encodings along the width axis is:

$$y_{ij} = \sum_{w=1}^{W} \mathrm{softmax}\left(q_{ij}^T k_{iw} + q_{ij}^T r_{iw}^q + k_{iw}^T r_{iw}^k\right)(v_{iw} + r_{iw}^v), \tag{3}$$

where $i$ and $j$ are the pixel positions along the width and height axes; and $r_{iw}^q$, $r_{iw}^k$, and $r_{iw}^v$ are relative position encodings (designed to preserve the location information along the width axis) regarding queries, keys, and values, respectively.[38] Note that the axial attention along the height axis is similar.

We set the number of heads to 8 for the multi-head axial attention layer. Using multi-head attention is like having multiple channels when we use convolution layers. We have eight sets of heads to capture different aspects of informative attention. Attention blocks have different depths for different levels, deeper attention blocks for higher feature levels. Specifically, the depths of attention blocks (number of axial attention layers in the block) from top to bottom are [2, 4, 6, 8, 10].

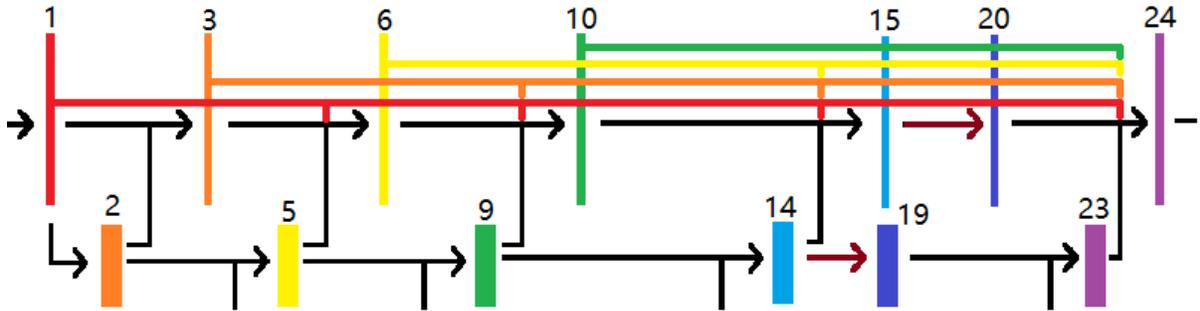

Figure 3. Skip-connection at the top level.

## 2.E. Feature map fusion and skip-connection

We introduce dense convolution layers and attention layers between the encoder and decoder paths. After each down-sampling and convolution step, we up-sample and fuse it with previously computed feature maps of larger scales until we restore the original resolution. Feature maps are fused by concatenation and a $1 \times 1$ convolution layer. We use dense skip-connections to preserve the semantics from the encoder-side feature maps and feed them to the decoder-side feature maps. The skip-connections of the top level are shown in Figure 3.

## 2.F. Experiments and evaluation metrics

We implemented the proposed WAU-net model in Python 3.7 and Pytorch and ran the in-house software on two NVIDIA 1080 Ti GPUs. We used the Adam optimizer[45] and a polynomial learning rate scheduler with the initial learning rate of 1e-3. The input images were center-cropped to $256 \times 256$ to remove the background without valuable information and to reduce the computation cost. We applied random shifting, rotation, and horizontal/vertical flipping to the input image and corresponding ground truth for data augmentation. We set the batch size to 4 and stopped the training when we see the model had converged. It took these models 500-700 epochs before convergence.

We evaluated the model performance on two datasets:

- The in-house dataset: we split the 115 cases into 90/25 for training and testing. With the 90 training cases, we used the 10-fold cross-validation strategy to select the best model and evaluated its performance on the test set.

- The Head and Neck 2015 challenge dataset: we performed the 5-fold cross-validation on the 33 training cases. We kept 3 randomly selected cases in the training set and split the rest 30 cases into 24/6 for each of the 5-fold cross-validation trials. After selecting the best model configuration, we trained our best model configuration with 30/3/15 train/validation/test split and used 15 testing cases to evaluate the model performance.

We used three commonly used evaluation metrics: Dice Similarity Coefficient (DSC), Hausdorff distance 95% (HD95), and mean surface distance (MSD).

- DSC[4] measures the spatial overlap between the prediction and ground truth:

$$DSC = \frac{2 \times |True \cap Pred|}{|True| + |Pred|}, \qquad (4)$$

where $|True|$ and $|Pred|$ are the number of pixels in the ground truth and prediction, respectively.

- The maximum Hausdorff distance is the maximum distance of a set $X$ to the nearest point in the other set $Y$, and the HD95[1] is based on the calculation of the 95$^{th}$ percentile of the distances between boundary points in $X$ and $Y$:

$$d_{HD95}(X,Y) = max_{95\%}(d_{XY}, d_{YX}) =$$
$$max_{95\%}[min_{y \in Y} d(x,y), max_{y \in Y} min_{x \in X} d(x,y)], \quad (5)$$

where $x$ represents a point on predicted contour $X$, and $y$ is a point on the ground truth contour $Y$; $d(x,y)$ is the Euclidean distance between point $x$ and point $y$.

- The MSD[1] measures the average distance between points on the predicted contours $X$ and the nearest point in ground truth contours $Y$:

$$MSD = \frac{1}{|X|+|Y|} (\sum_{x \in X} d(x,Y) + \sum_{y \in Y}(y,X)), \quad (6)$$

where $d(x,Y)$ is the distance between point $x$ to the closest point on $Y$.

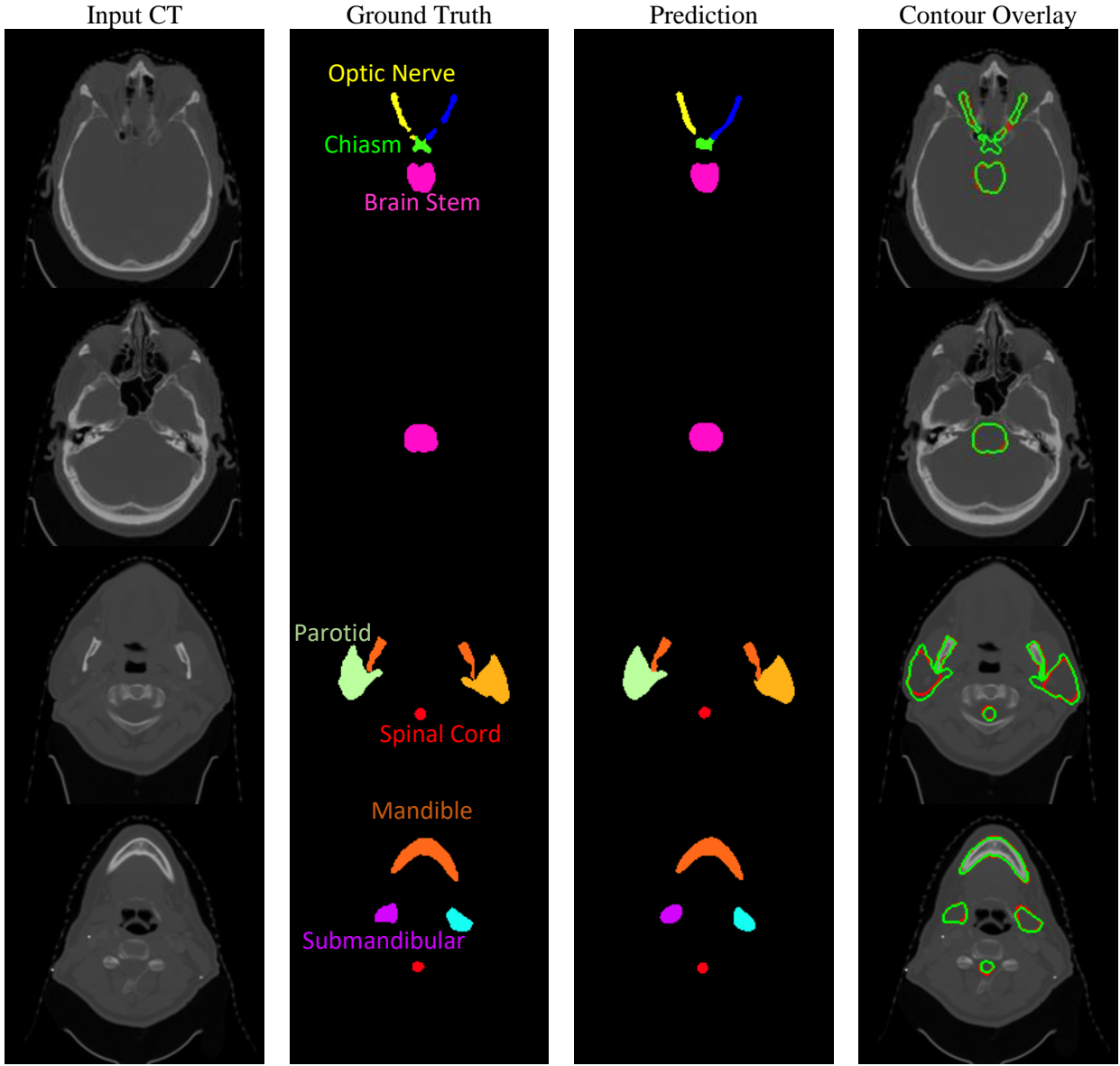

Figure 4. Segmentation result exhibition of a representative patient. The four slices are selected from inferior to superior. In the "Overlay" column, we use green lines or ground truth and red lines for prediction. OARs include brain stem, chiasm, mandible, spinal cord, left/right optic nerve, left/right submandibular, and left/right parotid.

## 3. RESULTS

### 3.A Results on the in-house dataset

The proposed WAU-net model generated contours that closely resemble the ground truth (Figure 4). The processing time for each patient is less than 2 minutes, which is a significant speed-up compared to manual delineation. For better visual comparison, we created "difference maps" (Figure 5) by coloring the overlap regions of ground truth and predictions in black and leaving the misclassified pixels in their original color. The comparison shows that the proposed model drastically outperformed the baselines.

| Input CT | Ground Truth | Prediction | Contour Overlay |
|---|---|---|---|
| U-net[27] | U-net++[28] | DAU-net[30] | WAU-net (ours) |
| 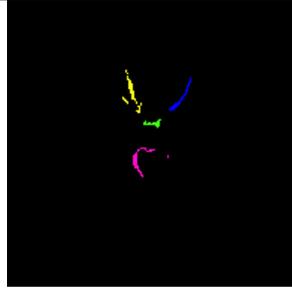 | 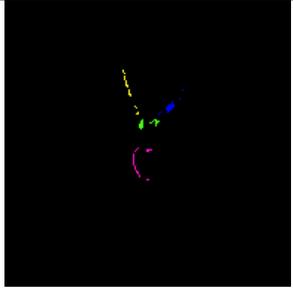 | 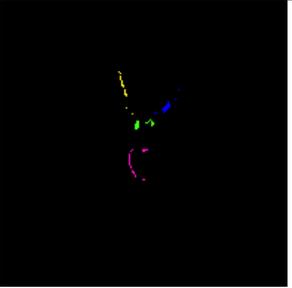 | 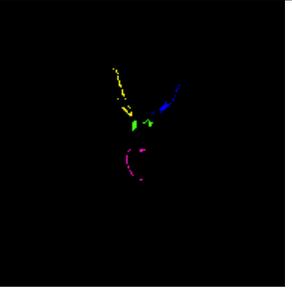 |
| 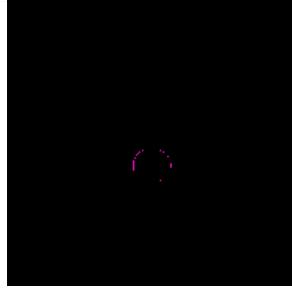 | 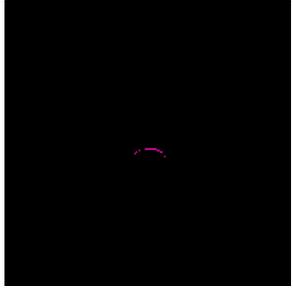 | 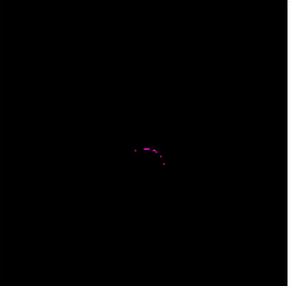 | 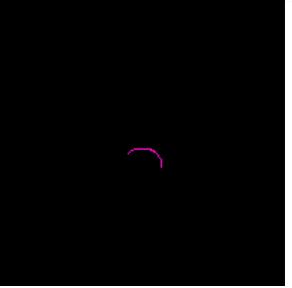 |
| 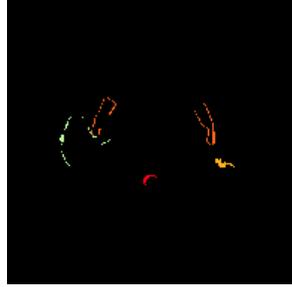 | 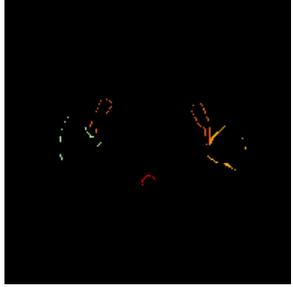 | 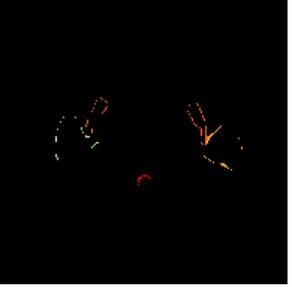 | 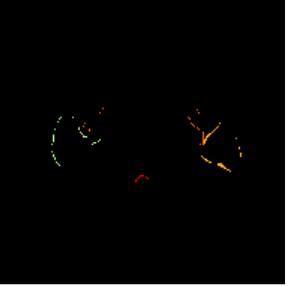 |
| 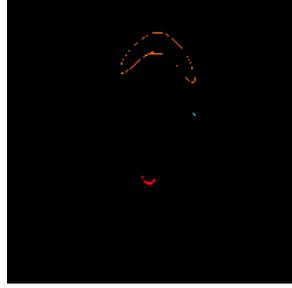 | 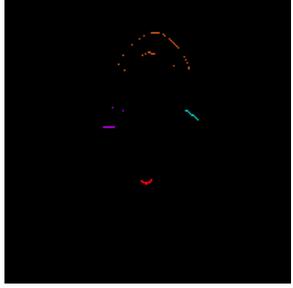 | 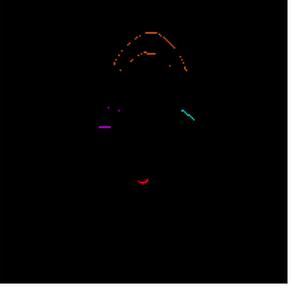 | 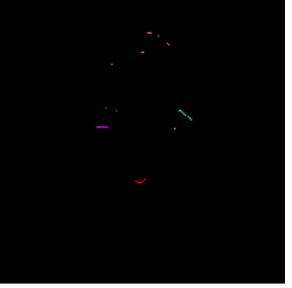 |

Figure 5. Difference map comparison of baseline methods and the proposed WAU-net.

We evaluated the model with the three performance metrics and compared it with baseline methods, including U-net[28], U-net++[29], and deep attention U-net (DAU-net[31]). Table I shows the comparison of our three metrics: DSC, HD95, and MSD. The results (Table II) showed that WAU-net significantly outperformed the baseline methods. The most improvement over the best baseline for each organ on the chiasm is 15.62%, and the least improvement on the spinal cord is 1.14%. The chiasm and optic nerves have the lower DSC due to their lower contrast and smaller size. They also have a higher performance increase in our proposed model because our multi-level attention mechanism and dense semantics forwarding captured and preserved the details of these finer organs.

| OAR | Metrics | U-net[27] | U-net++[28] | DAU-net[30] | WAU-net (ours) |
|---|---|---|---|---|---|
| Brain Stem | DSC | 0.84±0.04 | 0.90±0.05 | 0.88±0.05 | 0.92±0.03 |
| | HD95(mm) | 3.02 ± 1.24 | 2.11 ± 1.02 | 2.15 ± 1.01 | 1.95 ± 1.52 |
| | MSD | 1.21 ± 0.32 | 1.05 ± 0.28 | 1.02 ± 0.35 | 1.00 ± 0.24 |
| Chiasm | DSC | 0.42±0.15 | 0.55±0.14 | 0.64±0.08 | 0.74±0.18 |
| | HD95(mm) | 5.01 ± 2.55 | 4.17 ± 1.47 | 3.32 ± 1.46 | 2.42 ± 1.27 |
| | MSD | 1.61 ± 0.95 | 1.34 ± 0.79 | 1.52 ± 0.57 | 0.58 ± 0.46 |
| Mandible | DSC | 0.86±0.05 | 0.89±0.04 | 0.90±0.03 | 0.95±0.04 |
| | HD95(mm) | 2.45 ± 2.64 | 2.30 ± 1.89 | 2.42 ± 1.42 | 1.56 ± 1.89 |
| | MSD | 0.82 ± 0.41 | 0.80 ± 0.41 | 0.79 ± 0.47 | 0.64 ± 0.38 |
| Spinal Cord | DSC | 0.73±0.04 | 0.88±0.05 | 0.86±0.04 | 0.89±0.04 |
| | HD95(mm) | 3.22 ± 3.12 | 2.45 ± 2.00 | 2.04 ± 1.51 | 1.77 ± 1.78 |
| | MSD | 1.77 ± 0.82 | 1.52 ± 0.79 | 1.53 ± 0.38 | 1.24 ± 0.35 |
| Optic Nerve Left | DSC | 0.62±0.14 | 0.72±0.15 | 0.66±0.13 | 0.76±0.07 |
| | HD95(mm) | 8.88 ± 6.64 | 6.42 ± 5.21 | 8.12 ± 7.25 | 5.75 ± 5.14 |
| | MSD | 1.83 ± 1.75 | 1.75 ± 1.70 | 1.89 ± 1.51 | 1.68 ± 1.22 |
| Optic Nerve Right | DSC | 0.64±0.12 | 0.74±0.14 | 0.68±0.12 | 0.78±0.09 |
| | HD95(mm) | 8.24 ± 5.79 | 5.87 ± 4.75 | 8.20 ± 6.85 | 5.54 ± 5.16 |
| | MSD | 1.86 ± 1.78 | 1.76 ± 1.72 | 1.94 ± 1.60 | 1.66 ± 1.14 |
| Submandibular Left | DSC | 0.82±0.11 | 0.82±0.08 | 0.86±0.09 | 0.87±0.07 |
| | HD95(mm) | 3.41 ± 1.47 | 3.54 ± 1.25 | 3.22 ± 1.11 | 3.18 ± 1.21 |
| | MSD | 1.34 ± 0.65 | 1.35 ± 0.62 | 1.22 ± 0.51 | 1.21 ± 0.48 |
| Submandibular Right | DSC | 0.85±0.10 | 0.83±0.10 | 0.84±0.09 | 0.85±0.09 |
| | HD95(mm) | 3.23 ± 1.70 | 3.44 ± 1.76 | 3.34 ± 1.83 | 3.01 ± 1.45 |
| | MSD | 1.02 ± 0.52 | 1.12 ± 0.54 | 1.12 ± 0.45 | 1.16 ± 0.56 |
| Parotid Left | DSC | 0.83±0.05 | 0.86±0.08 | 0.85±0.05 | 0.86±0.06 |
| | HD95(mm) | 3.11 ± 0.74 | 3.01 ± 1.29 | 3.05 ± 1.21 | 2.98 ± 0.74 |
| | MSD | 1.12 ± 0.25 | 1.04 ± 0.27 | 1.10 ± 0.24 | 1.08 ± 0.32 |
| Parotid Right | DSC | 0.85±0.06 | 0.86±0.05 | 0.86±0.05 | 0.88±0.05 |
| | HD95(mm) | 3.52 ± 0.88 | 3.24 ± 1.44 | 3.45 ± 1.62 | 3.11 ± 1.01 |
| | MSD | 1.17 ± 0.29 | 1.09 ± 0.30 | 1.17 ± 0.39 | 1.12 ± 0.38 |

Table I. Mean and standard deviation of the three metrics (DSC, HD95, and MSD) for our WAU-net and baseline methods. Higher DSC, lower HD95, and lower MSD indicate better overlapping between predictions and ground truths.

| | U-net[27] | U-net++[28] | DAU-net[30] |
|---|---|---|---|
| p-value | 0.01 | 0.02 | 0.02 |

Table II. p-value between three baseline methods and our WAU-net.

### 3.B Results on the head and neck 2015 challenge data

The Head and Neck 2015 challenge dataset provides delineation ground truths for nine OARs: brain stem, chiasm, mandible, left/right optic nerve, left/right parotid, and left/right submandibular. The results (Table III) showed that our WAU-net's overall performance is comparable to state-of-the-art methods. WAU-net achieved the highest performance on the submandibular and the right optic nerve, and for the rest of OARs, it had similar scores to the existing state-of-the-art methods.

| OAR | 2-stage 3D U-net[21] | Shape Representation Model + FCN[23] | UaNet[33] | Anatomy Net[32] | Dual Pyramid Network[1] | WAU-net (challenge) | WAU-net (in-house) |
|---|---|---|---|---|---|---|---|
| Brain Stem | 0.88±0.02 | 0.87±0.03 | 0.87±0.03 | 0.87±0.02 | 0.91±0.02 | 0.91±0.02 | 0.84±0.03 |
| Chiasm | 0.45±0.17 | 0.58±0.1 | 0.62±0.1 | 0.53±0.15 | 0.73±0.11 | 0.71±0.17 | 0.66±0.15 |
| Mandible | 0.93±0.02 | 0.87±0.03 | 0.95±0.01 | 0.93±0.02 | 0.96±0.01 | 0.95±0.03 | 0.90±0.04 |
| Optic Nerve Left | 0.74±0.15 | 0.65±0.05 | 0.75±0.07 | 0.72±0.06 | 0.78±0.09 | 0.76±0.08 | 0.67±0.09 |
| Optic Nerve Right | 0.74±0.09 | 0.69±0.5 | 0.72±0.06 | 0.71±0.1 | 0.78±0.11 | 0.79±0.05 | 0.71±0.05 |
| Submandibular Left | 0.76±0.15 | 0.76±0.06 | 0.82±0.05 | 0.81±0.04 | 0.86±0.08 | 0.87±0.05 | 0.82±0.05 |
| Submandibular Right | 0.73±0.01 | 0.81±0.06 | 0.82±0.05 | 0.81±0.04 | 0.85±0.10 | 0.86±0.08 | 0.81±0.06 |
| Parotid Left | 0.86±0.02 | 0.84±0.02 | 0.89±0.02 | 0.88±0.02 | 0.88±0.04 | 0.87±0.03 | 0.82±0.04 |
| Parotid Right | 0.85±0.07 | 0.83±0.02 | 0.88±0.05 | 0.87±0.04 | 0.88±0.06 | 0.87±0.07 | 0.83±0.06 |

Table III. The means and standard deviations of DSC for WAU-net and the state-of-the-art methods. We also include the model trained with our in-house data to evaluate the generalizability.

## 4. DISCUSSION

The proposed WAU-net method effectively integrated the self-attention mechanism into a CNN architecture and overcame the weakness of CNNs in long-range dependency modeling. Introducing axial attention layers into the CNN architecture enabled explicit modeling of pixel-to-pixel relations, resulting in finer segmentation outcomes for OARs at different scales.

Three representative algorithms were selected as the baselines for comparison. The first baseline, U-net[28], is one of the classic architectures for segmentation and has been broadly adopted for medical image segmentation. The encoder-decoder design inspired many works, among which we selected U-net++[29] as the second baseline because it has a similar forward passing path design built with only CNNs. WAU-net outperformed U-net++, indicating the effect of the multi-level attention mechanism used. The third baseline, DAU-net[31], is proposed to integrate feature maps and attention-aware techniques and make the upsampling process in the decoder trainable and meaningful. The attention-aware fusion block in DAU-net uses $1 \times 1$ convolution layers to map input feature maps to queries and keys, and it does not involve value matrices, making it fundamentally different and weaker compared to the axial attention block that we used in WAU-net. The superior performance of WAU-net over DAU-net demonstrated the effectiveness of multi-scale pixel-to-pixel attention blocks and dense forwarding paths for preserving semantics.

We compared WAU-net to the state-of-the-art methods on the Head and Neck Challenge 2015 dataset. Our method achieved the best or the second-best performance on all OARs. For each OAR, the 2-stage 3D U-net[22] uses a 2D U-net localization network to extract image patches and a 3D U-net segmentation network to generate contours, giving rise to 18 networks in total. Each OAR needs to be segmented separately and then fused into the final segmentation map. The shape representation plus FCN method[24] combines the merits of the traditional shape-representation model and the CNN-based methods. However, it implements a shape representation encoder-decoder network to reconstruct the CT images before feeding into the segmentation FCN. Compared to these two methods, a major advantage of our method is that we only need to train one network, and this trained network can segment nine OARs within a single forward propagation. Our method significantly improves inference efficiency. The UaNet[34] and AnatomyNet[33] adopt the U-net structure and implement different design upgrades. They have good performance while still suffering from the weakness of CNNs. Our method outperformed these two architectures by introducing the self-attention mechanism. Dual Pyramid Network[1] leverages the power of synthetic MRIs generated by a pre-trained Cycle-GAN. It synthesizes an MRI for each input CT image and used two DAU-nets to predict the segmentation label maps. The method achieved the state-of-the-art performance on multiple organs by capturing the finer details of organs that have higher contrast on MRIs compared to CTs. Our method achieved comparable performance but did not need a pre-trained Cycle-GAN, whose training requires large datasets, high computation costs, and a delicate tuning process.

Consdiering the generalization ability of the model, we used the WAU-net trained with our in-house dataset to draw contours for the challenge dataset. We discarded the contour for the spinal cord to be consistent with the challenge dataset. The results showed that our model maintains great performance on all OARs (Table III) thanks to its generalization ability and proper data augmentation during the training phase, and the performance loss is due to differnet data distributions between the two datasets.

The limitations of this work are two-fold. First, it is computationally expensive to use dense CNN and attention connections between the encoder and decoder path and compute pixel-to-pixel self-attention, which is unscalable for high-resolution images. Secondly, the datasets used in the current study are limited. We will test the model performance on larger datasets from different hospitals to make the model robust.

## 5. CONCLUSION

We presented a novel deep learning method for semantic segmentation of head and neck CT images. The new WAU-net method integrated the axial attention layers with CNN frameworks to efficiently capture local and global dependencies. With multi-scale axial attention layers and dense semantics preserving paths, we refined the segmentation outcomes for larger and finer OARs. Evaluated on an in-house dataset and the Head and Neck 2015 Challenge dataset, WAU-net outperformed all baseline methods on the in-house data and achieved better or comparable performance with the baseline methods on the challenge dataset. The new method showed promising clinical value by significantly speed-up the delineation process and providing accurate OARs contours.

## ACKNOWLEDGEMENTS

We thank Varian Medical System for their financial support through a research grant.